\documentclass[12 pt,english]{iopart}
\usepackage{graphicx}
\usepackage{graphics}

\begin{document}

\title{Loop variables in the geometry of a rotating black string}

\author{A. M. de M. Carvalho and Fernando Moraes }

\address{ Laborat\'orio de F\'{\i}sica Te\'orica e Computacional,\\ 
Departamento de F\'{\i}sica\\ 
Universidade Federal de Pernambuco, 50670-901, Recife, PE, Brazil}
\author{ Claudio Furtado}
\address{ Departamento de F\'{\i}sica,\\ Universidade Federal da Para\'{\i}ba,Caixa Postal 5008, 58051-970, Jo\~ao Pessoa, PB, Brazil}
\begin{abstract}
In this paper we analyze in the Wilson loop context the parallel 
transport of vectors and spinors around a closed loop
in the background space-time of a rotating black string 
in order to classify its global properties. We also examine
particular closed orbits in this space-time and verify the Mandelstam
relations.
\end{abstract}
\pacs{04.20-q, 04.70.Bw, 04.20.Cv }

\submitto{\CQG}

\section{Introduction}


 In the early sixties Mandelstam \cite{anp:m} proposed a new formalism for
electrodynamics and gravitation in which the fields depend on the paths 
rather than on space-time points. The fundamental quantity that arises 
from this path-dependent approach, the non-integrable phase factor\cite{prd:wu}
(loop variable) represents the field more adequately than the field strength
does. In the application of this formalism in the theory of gravity,
Mandelstam\cite{anp:m} established several equations involving the loop 
variables. Some years later Voronov and Makeenko\cite{sjnp:vm} showed the 
equivalence between the equations 
obtained using this approach and Einstein's equations.

The quantities we intend to use are the gravitational loop variables and 
corresponding holonomies of the Christoffel connection or of the spin 
connection. Keeping the local fields as the fundamental dynamical variables,
we shall compute the loop variables in different cases in order to learn 
about their behavior and their geometrical meaning. 

The loop variables in the theory of gravity are matrices representing parallel
transport along paths in a space-time with a given affine connection. They 
are defined as the limit of an ordered product of matrices of infinitesimal 
parallel transport as

\begin{equation}
U_\nu ^\mu (C_{yx};\Gamma )\equiv \prod\limits_{i=1}^N(\delta _{\rho
_{1-i}}^{\rho _i}-\Gamma _{\lambda _i\rho _{1-i}}^{\rho
_i}(x_i)dx_i^{\lambda _i})  \label{eq1},
\end{equation}
where $x_0=x,$ $\rho _o=\nu ,$ $x_N=y,$ $\rho _N=\mu ,$ $dx_i=(x_i-x_{i-1})/%
\varepsilon$.

The points $x_i$ lie on an oriented curve $C_{yx}$ with its beginning at the
point $x$ and its end at the point $y$. The parallel-transport matrix $%
U_{\nu ^{}}^\mu $ is a functional of the curve $C_{yx}$ as a geometrical
object.

If we choose as tetrad frame, a basis $\left\{ e_b^a(x)\right\} $ and a
loop $C$ such that $C(0)=C(1)=x$, then in parallel transporting a vector $
X^\alpha $ from $C(\lambda )$ to $C(\lambda +d\lambda )$, the vector
components change by 
\begin{equation}
\delta X^\mu =M_\nu ^\mu \left[ x(\lambda )\right] X^\nu d\lambda ,
\label{eq2}
\end{equation}
where $M _{\nu}^\mu $ is an infinitesimal linear map which depends
on the tetrads, on the affine connection of the space-time and on the value
of $\lambda .$ Therefore, the linear map takes any tangent vector to $C$ in a 
given point and parallel transport it along $C$ back to the initial point.
Then, it follows that the holonomy transformation $U_\nu ^\mu $
is given by the matrix product of the $N$ linear maps\thinspace 
\begin{equation}
U_\nu ^\mu =\lim\limits_{N\rightarrow \infty }\prod\limits_{i=1}^\infty
\left[ \delta _\nu ^\mu +\frac 1NM_\nu ^\mu \left[ x(\lambda )\right]
|_{\lambda =\frac iN}\right]  \label{eq3}.
\end{equation}
One often writes the linear map $U_\nu ^\mu $ given by Eq.(\ref{eq3}) as 
\begin{equation}
U(C)={\cal P}\exp \left( \int\limits_CM\right) ,  \label{eq4}
\end{equation}
where ${\cal P}$ means ordered product along a curve $C$. Equation (\ref{eq4})
should be understood as an abbreviation of the right hand side of 
eq.(\ref{eq3}). Note that if $M_\nu ^\mu $ is independent of $\lambda ,$ 
then it follows from eq.(\ref{eq3}) that $U_\nu ^\mu $ is given by 
$U_\nu ^\mu =\left( \exp M\right)_\nu ^\mu$. The linear map $U_\nu ^\mu$
is called holonomy transformation associated with the curve $C$. From
eq.(\ref{eq4}) we can construct the invariants which involve the holonomy 
group. This group measures the deviation of the space from global flatness. 
The basic invariant associated with $U(C)$ is the trace of $U(C)$. This 
quantity provides information about the geometric and physical structure of 
the space-time.
In this paper we shall use the following notation
\begin{equation}
U_{AB}(C)={\cal P}\exp \left(- \int\limits_B^A\Gamma _\mu (x(\lambda ))\frac{dx^\mu }
{d\lambda }d\lambda \right) ,  \label{eq6}
\end{equation}
where $\Gamma _{\mu \textrm{ }}$ is the tetradic connection and $A$ and $B$
are the initial and final points of the path. Then, associated with every
path $C$ from a point $A$ to a point $B$, we have a loop variable $U_{AB}$
given by eq.(\ref{eq6}) which, by construction, is a function of the path $C$
as a geometrical object.
Using holonomy, Bollini, Giambiagi and Tiomno\cite{lanc:bgt} 
investigated the Kerr
black holes space-times and obtained many properties of this geometry.
In a recent article, Rothman, Ellis and Murugan \cite{cqg:rem}
investigated the holonomy in the Drost-Schwarzschild geometry
obtaining very interesting results for a class orbits
in this space-time. They found that holonomy in this space-time 
has a ``quantization''  property denominated band invariance holonomy. 
Those results can be used for the investigation of strong-gravity 
effects~\cite{cqg:abramo} in the X-ray emission from a few
acreting galactic black holes and neutron star.
Their results can be used for the study
of the Einstein-Podolski-Rosen experiment in a curved background~\cite{pla:mens}.
In a recent article the clock effects were investigated by Bini, Jantzen 
and Mashhoon\cite{cqg:bjm}, via holonomy transformation. 
Burgers\cite{prd:b} and Bezerra\cite{prd:v} examined the effects
of a parallel
transport of vectors and spinors both around a point-like
solution and a cylindrically symmetric cosmic string.
This procedure  gives, in general, non-trivial results.
These effects point out to the gravitational analogue
Aharonov-Bohm effect. Two of us (CF and FM)\cite{gc:afbm} investigated
by holonomy transformation, the topological properties of
a class of solutions in Kaluza-Klein theory and demonstrated that
the holonomy gives a combined effect of gravitation and electromagnetic
fields in the parallel transport of vectors in that space-time.
This combined the electromagnetic and gravitational
Aharonov-Bohm effects. The analysis of Berry's
quantum phase in gravitation\cite{prd:claudassis} was studied in a recent article.
The use of holonomy for quantum computation  was analyzed
in a geometric approach by Pachos and Zanardi\cite{ijmp:pz}.


\section{The Geometry of a Black String}


Black holes constitute one of the most fascinating structures
of nature. They have been extensively studied in a wide variety of
models. Recently, Ba\~nados, Teitelboim and Zanelli\cite{prl:btz, prd:bhtz}
obtained a $(2+1)$-dimensional solution of Einstein's field equations
with negative cosmological constant, called BTZ black hole. Several authors 
have used gauge formulation of gravity in (2+1) dimensions for a description 
the BTZ solution\cite{prd:can,plb:witt}, they have demonstrated that Wilson 
loops in the three dimensional  Anti-de Sitter group SO(2,2) reproduce the
 BTZ geometry.  This solution
is very important to the study of many aspects of black hole quantum physics,
thermodynamics and other general properties. String-like configuration of 
black holes have been studied in recent years\cite{npb:horne,prd:kalo}, 
this object can be seen with a foliation of tridimensional objects. 
More recently,
Lemos and Zanchin\cite{prd:lz} transformed the $(2+1)$ BTZ black hole into
a $(3+1)$ cylindrical black hole, known as a black string. This process is
similar to the transformation of a punctual particle in $(2+1)$
dimensions into a straight cosmic string in four-dimensional space-time. 
This spacetime is asymptotically Anti-de Sitter in the radial direction 
and non-asymptotically flat. Its geometric structure is that of the BTZ 
black hole extended with a flat line which is interpreted as the string axis. 
Its asymptotic symmetry group is ${\bf R}$ times the conformal group 
in two dimensions of which ${\bf R\times SO(2,2)}$ is a subgroup.

We intend to use the gravitational
loop variable and corresponding holonomy of the spin connection  to
describe the geometry of the space-time of a black string. The study 
of this holonomy permits us to investigate the global
properties of this space-time. The aim of this paper is 
to investigate the holonomy for
a class of orbits in the black string geometry.

The line element of a black string space-time may be written
as \cite{prd:lz}
\begin{equation}
\label{btz}
ds^{2}=-N^{2}dt^{2}+N^{-2}dr^{2}+r^{2}\left(N^{\phi}dt+
d\phi \right)^{2}+dz^{2},
\end{equation}
$N$ and $N^{\phi}$ are called
lapse and angular shift functions and are given  respectively
by

\begin{eqnarray}
N^{2}(r) &=& -8M+\alpha^{2}r^{2}+\frac{16J^{2}}{r^{2}}\mbox{,}   \\
N^{\phi}(r) &=& -\frac{4J}{r^{2}} \mbox{,}
\end{eqnarray}
where $M$ is a term associated with the mass, $J$, the angular momentum
term and $\alpha$ is associated with the cosmological constant,  $\Lambda$,
more precisely,  $\Lambda=-\alpha^{2}$. The non-rotational case
is gotten when we make the angular momentum null, $J=0$.
Introducing an appropriate dual $1$-form basis(co-frame), defined 
by $e^{a}=e^{a}_{\mu}dx^{\mu}$,
where
\begin{eqnarray}
        e^{0}&=& Ndt\mbox{,}       \\
        e^{1}&=& N^{-1}dr\mbox{,}  \\
        e^{2}&=& r(N^{\phi}dt +d\phi)\mbox{,}    \\
        e^{3}&=& dz\mbox{,}
\end{eqnarray}
the 1-form connections are obtained from the first of the
Maurer-Cartan structure equations: 
$de^{a}+\omega ^{a}_{b} \wedge e^{b}=0$.
 From the above basis we find the following connections:
\begin{eqnarray}
\omega^{0}_{1}&=& \omega^{1}_{0}= N N' dt \mbox{,}       \\
\omega^{1}_{2}&=& -\omega^{2}_{1}= -N(N^{\phi}dt +d\phi) \mbox{.}
\end{eqnarray}
This result takes us to the following spin connection
for the black string metric
\begin{eqnarray}
\Gamma_{\phi} =\left( \begin{array}{cccc}
0 & 0 & 0 & 0  \\ 
0 & 0 & -N & 0 \\ 
0 & N & 0 & 0  \\
0 & 0 & 0 & 0
\end{array}\right) .
\end{eqnarray}
This result is the same obtained when we consider a non-rotating black string.
Also 
\begin{eqnarray}
\Gamma_{t} =\left( \begin{array}{cccc}
0 & N N'  & 0 & 0  \\ 
N N'  & 0 & -NN^{\phi} & 0  \\
0 & NN^{\phi} & 0 & 0  \\ 
0 & 0 & 0 & 0
\end{array}\right) .
\end{eqnarray}
When we make $N^{\phi}=0$, this result is reduced to the non-rotating
black string case.
The spin connection $\Gamma_{\phi}$ corresponds to closed
curves with constant time, while $\Gamma_{t}$ corresponds
to closed curves with constant azimuthal angle.
The holonomy matrix associated with the
parallel transport of vectors around closed curves $\gamma$
is defined by
\begin{equation}
\label{holonomia}
U(\gamma)={\cal P}\exp\left(-\oint _{\gamma} \Gamma_{\mu}dx^{\mu}\right),
\end{equation}
where the ${\cal P}$ is the order operator.

We are interested in analyzing closed paths around a black string.
First, we will study closed curves with
constant time and centered  at the origin.
As $\Gamma_{\phi}$ is independent of the azimuthal angle $\phi$, then
the holonomy integral (\ref{holonomia}), can easily be calculated
\begin{eqnarray}
\label{h-phi}
U(\gamma)&=&\exp \left(-\oint \Gamma_{\phi}d\phi \right), \nonumber \\
&=&\exp\left(-2\pi\Gamma_{\phi}\right).
\end{eqnarray}
Making the expansion of this equation and noticing that  we are always
able to write the exponents of upper order in 
$\Gamma_{\phi}$ and $\Gamma_{\phi}^{2}$ terms.
Then eq. (\ref{h-phi}) can be written in compact form
in the following way:
\begin{equation}U(\gamma)= 1- \frac{\Gamma_{\phi}}{N}\sin(2\pi N)+
\frac{\Gamma_{\phi}^{2}}{N^2}\left[1-\cos(2\pi N)\right].
\end{equation}
The above expression corresponds to the holonomy transformation
for a circular path with constant time.

We observe that these holonomy transformations have the same property 
analyzed by Ellis and coworkers in Schwarchild case, the band holonomy 
invariance. Note that after n loops in the afore mentioned path we observe 
that if  $2\pi n N =2\pi m$  there exists a critical radius, $r_{cr}$,
given by:
\begin{equation}
R^{2}=\frac{1}{2\alpha^{2}}\left(8M+(m/n)^{2}\pm \sqrt{(8M+(m/n)^{2})^{2}-(8\alpha J)^{2}}  \right),
\end{equation}
where the holonomy is trivial and we obtain a behavior similar to the 
observed in Schwarzschild case. For this case the holonomy matrix is trivial 
and the transported vector does not acquire a deficit angle in this transport. 

Writing the holonomy in a matrix form, we set
\begin{eqnarray}
U(\gamma)=\left(
\begin{array}{cccc}
1 & 0 & 0 & 0\\
0 & \cos(2\pi N) & \sin(2\pi N) & 0 \\
0 & -\sin(2\pi N) &  \cos(2\pi N) & 0 \\
0 & 0 & 0 & 1
\end{array}
\right) .
\end{eqnarray}
This matrix can be interpreted as the rotation generator
around the $z$-axis, since $U(\gamma)=\exp(-2\pi i N J_{12})$.
We can get a quantization condition by imposing that $nN=m$,
where $m$ is an integer. For these values the holonomy matrix is 
equal the  identity matrix.

The deficit angle $\chi$ obtained when we compare the final and initial
position of the parallel transported vector is given by
\begin{equation}
\cos \chi_{A}=U_{A}^{A},
\end{equation}
where $A$ is a tetradic index. The terms of non vanishing angular
deviations occur when $A=1$ and $2$, so we have
\begin{eqnarray}
\cos\chi_{1  or  2}=\cos(2 \pi N)
\end{eqnarray}
or
\begin{eqnarray}
|\chi_{1  or  2}|=|2 \pi N + 2\pi n|.
\end{eqnarray}
For $N \rightarrow 0$ we must have $\chi_{1 or 2} \rightarrow 0$,
so we choose $n=0$, which leads to
\begin{eqnarray}
\label{angu}
|\chi_{1  or  2}|=|2\pi N|.
\end{eqnarray}
The above expression shows that for $N\neq 0$, if we parallel
transport a vector around a closed path the final vector 
does not coincide with the original vector. This physical effect
could be understood as a gravitational analogue of 
the Aharonov-Bohm effect. We saw in (\ref{angu}) that there will 
be no Aharonov-Bohm effect if $2\pi N$ is an integer. This conditions is 
not always satisfied, because $N$ is not necessarily an integer. In fact, 
it can assume an arbitrary value. There are  similarities of this effect 
with the Aharonov-Bohm effect in the cosmic string case~\cite{prd:v}. 
The crucial difference occurs in the fact that the spacetime exterior to 
the cosmic string has Riemann tensor null  contrary to the black 
string case that has constant curvature in the exterior region.

The natural way to construct an invariant quantity 
under a gauge transformation  is determining the
Wilson loop. The corresponding 
Wilson loop for a closed path $\gamma$  is defined by
\begin{equation}
W(\gamma)=tr {\cal P} \exp \left[ -\oint_{\gamma} dx^{\mu}\Gamma_{\mu}(x) \right].
\end{equation}
Ashtekar~\cite{prl:ashtekar} ``rebuild'' general relativity
as a ``loop representation''. In this formalism
the fundamental variables are the holonomies instead
of the metrics.
This approach allowed advances in the direction of a 
quantum theory of gravity.
For a closed path with constant time the Wilson loop is given by
\begin{equation}
W(\gamma)=2\left( 1 + \cos 2\pi N \right).
\end{equation}
\noindent
Note that the Wilson loop is trivial for the same values were occurs
the band holonomy invariance $2\pi n N=2 \pi m$. In this way, we can use 
the Wilson loop to investigate the band holonomy invariance.  

Another path can be obtained by fixing
the azimuthal angle,
\begin{eqnarray}
\label{h-t}
U(\gamma)&=&\exp \left(-\oint \Gamma_{t}dt \right) \nonumber \\
&=&\exp\left(-T\Gamma_{t}\right),
\end{eqnarray}
which leads to the following expression for
the phase (\ref{holonomia})
\begin{eqnarray}
U(\gamma)= 1- \frac{\Gamma_{t}}{\sqrt{(N N')^{2}-(NN^{\phi})^{2}}}
\sinh(\sqrt{(N N')^{2}-(NN^{\phi})^{2}}T)+ \nonumber \\
\frac{\Gamma_{t}^{2}}{(N N')^{2}-(NN^{\phi})^{2}}\left[\cosh(\sqrt{(N N')^{2}-(NN^{\phi})^{2}}T)-1\right].
\end{eqnarray}
In this case, we note that the holonomy carries information
about the term associated with the cosmological constant.
Now writing $U(\gamma)$ in its matrix form we obtain,
\begin{eqnarray}
U(\gamma)=\nonumber \\ 
\left(
\begin{array}{cccc}
\cosh(\sqrt{(N N')^{2}-(NN^{\phi})^{2}}T)  & -\sinh(\sqrt{(N N')^{2}-(NN^{\phi})^{2}}T) & 0 & 0 \\
-\sinh(\sqrt{(N N')^{2}-(NN^{\phi})^{2}}T) & \cosh(\sqrt{(N N')^{2}-(NN^{\phi})^{2}}T) & 0 & 0\\
0 & 0 & 1 & 0 \\
  0 & 0 & 0 & 1\end{array}\right) .
\end{eqnarray}
We identify $ (N N')^{2}-(NN^{\phi})^{2})T $ 
as the ``boost" parameter
in a Lorentz transformation. The Wilson loop for a closed path
with constant azimuthal angle is given by
\begin{equation}
W(\gamma)=2\left( 1 + \cosh(\sqrt{(N N')^{2}-(NN^{\phi})^{2}}T)  \right).
\end{equation}

The Wilson loop has drawn considerable attention because it is a gauge-invariant quantity 
and because it can act as a dynamical variable. It is therefore interesting to obtain the 
Wilson loop and to verify the relations it obeys for the case studied here.
Mandelstam has proved that the loop variables satisfies a series of relations.
Our  interest is now in the proof of this relation for the black string case.
Now, we use the curvature $2$-form,
$R_{b}^{a}$, in order to verify the Mandelstam relations.
$R_{b}^{a}=d\omega_{b}^{a}+\omega_{c}^{a} \wedge \omega_{b}^{c}.$
This equation is called the second Maurer-Cartan structure equation.
In this way the curvature matrix for a constant time curve can
be evaluated as
\begin{eqnarray}
R_{12} =\left(
\begin{array}{cccc}
        0 & 0 & 0 & 0\\
        0 & 0 & N' & 0 \\
        0 & -N' & 0 & 0 \\
        0 & 0 & 0 & 0
\end{array}\right) ,
\end{eqnarray}
where the prime denotes the derivative  with respect to the
variable $r$. The Mandelstam relation is given by
\begin{equation}
\frac{\partial W}{\partial x_{\nu}}=
        \oint dS Tr \left[R_{\nu \mu}U(\gamma)\right]
        \frac{dy^{\mu}}{dS}.
\end{equation}
This relation appears in the quantum gravitation loop
variable theory.  But,
\begin{equation}
\frac{\partial W}{\partial r}=-4\pi N'\sin(2\pi N),
\end{equation}
then, it follows that
\begin{equation}
\frac{\partial W}{\partial r}=\int Tr [R_{21}U(\gamma)] d\phi.
\end{equation}
Notice that the Mandelstam relations  must 
be satisfied by the loop variables in a nonperturbative approach to the 
quantization of gravity \cite{anp:m}. In this work 
we have shown that they are obeyed classically as well.

Now we consider a more general path in this geometry that performs a helix 
in space-time. In this trajectory only the radial coordinate is 
maintained constant.
We choose a more general path now, where 
$\Gamma_{\mu}=\Gamma_{t}dt+\Gamma_{\phi}d\phi$,
In this case the holonomy is given by
\begin{equation}
U_{G}(\gamma)=\exp \left(- \int_{0}^{2\pi n}
(\beta\Gamma_{t}+\Gamma_{\phi}) \right)d\phi.
\end{equation}
Expanding the above exponential, we get
\begin{equation}
U_{G}(\gamma)=1-2\pi r \Gamma_{s}+ \frac{(2\pi\Gamma_{s})^{2}}{2!}
+\frac{(2\pi\Gamma_{s})^{3}}{3!} +\cdot\cdot\cdot ,
\end{equation} 
where $\Gamma_{s}=\beta\Gamma_{t}+\Gamma_{\phi}$.
The matrix form of $\Gamma_{s}$ is given by
\begin{eqnarray}
\Gamma_{s}
=\left(
\begin{array}{cccc}
0          & \beta N N'         & 0                    & 0\\
\beta N N' & 0                  & -\beta(NN^{\phi}+N)  & 0 \\
0          & \beta(NN^{\phi}+N) & 0                    & 0 \\
0          & 0                  & 0                    & 0
\end{array}
\right),
\end{eqnarray}
this matrix possesses the following properties:
$\Gamma_{s}^{3}=-\omega^{2}\Gamma_{s}$,
$\Gamma_{s}^{4}=-\omega^{2}\Gamma_{s}^{2}$,
and thus successively, where 
$\omega^{2}=N^{2}(N^{\phi}\beta+1)^{2}-\beta^{2}(NN')^{2}$.
The loop variable for general trajectories encircling the defects can 
be written in the following form
\begin{eqnarray}
\label{gerador}
U=\exp\left\{i \int_{0}^{2\pi n}(\beta NN'J_{01} + \beta N(N^{\phi}+1)J_{12})d\phi \right \} ,
\end{eqnarray}
where $J_{01}$ and $J_{12}$ are the generators of the $R\times SO(2,2)$ group, 
in this way it is possible to describe the holonomy in this space-time in the 
general way in function of generators of the symmetry group
\begin{eqnarray}
\label{geral holo}
U(\gamma)={\cal P}\exp\{ -i\int_{c}\Gamma_{\mu}^{ab}J_{ab}dx^{\mu}\}.
\end{eqnarray}

We return to the matrix $U(\gamma)$, regrouping the terms of the exponents 
odd and evens we can write the general holonomy as being 
\begin{equation}
U_{G}(\gamma)=1-\frac{\Gamma_{s}}{\omega}\sin(2\pi n\omega)
+\frac{\Gamma_{s}^{2}}{\omega^{2}}\left(\cos (2\pi n \omega)-1\right),
\end{equation}
were we consider that $0<\phi<2 \pi n$. In this way the band 
invariance holonomy can argued for some special orbit. Since 
$\omega > 0$ we have oscillatory solutions. In the next section 
we use the transport equation to examine this special orbit.


\section{Study of some Special Orbits}


Recently, Rothman, Ellis and Murugan \cite{cqg:rem} studied
some particular curves when a vector ${\bf A}$ is parallel
transported in the geometry of Reissner-Nordstr\"on and then observed 
very interesting properties of parallel transport that they denominated 
invariance band holonomy. In this section we analyze the band holonomy 
properties from 
the point of view of the parallel transport equation in the background 
of the rotating black string. First we consider the general case of a circular 
orbit, and then we study the constant time orbit and circular geodesics.
The components of the vector ${\bf A}$ satisfy the equation
\begin{equation}dA^{\mu}+\omega^{\mu}_{\beta}A^{\beta}=0,
\end{equation}
which is  the parallel transport equation for the
connection $\omega^{\mu}_{\beta}$. For the black string
metric (\ref{btz}), the components of the vector obey
the following equations
\begin{eqnarray}
\label{comp1}
dA^{0}+N N'dtA^{1}= 0  \mbox{,}      \\
\label{comp2}
dA^{1}+N N'dt A^{0}-N(N^{\phi}dt +d\phi)A^{2} = 0  \mbox{,} \\
dA^{2}+N(N^{\phi}dt +d\phi)A^{1}= 0  \mbox{,}         \\
dA^{3}=0 \mbox{.}
\end{eqnarray}
Now we study a family of particular solutions of the above differential 
equations system. The first particular solution
corresponds to  \underline{circular orbits}, $r=R$, so we
have a tangent vector like $X^{\mu}=(X^{t},0,X^{\phi},0)$.
In addition to this we parametrize the curve assuming that
the `velocity' is constant; i.e, $\beta=dt/d\phi$, so
we find
\begin{eqnarray}
\frac{dA^{0}}{d\phi}+\beta N N' A^{1}= 0  \mbox{,}       \\
     \frac{dA^{1}}{d\phi}+\beta N N' A^{0}-N(N^{\phi}\beta +1)A^{2}  = 0  \mbox{,}        \\
     \frac{dA^{2}}{d\phi}+ N(N^{\phi}\beta +1)A^{1}= 0  \mbox{,}
     \\      \frac{dA^{3}}{d\phi}=0 \mbox{.}
\end{eqnarray}
These equations can be easily integrated to
\begin{eqnarray}
A^{0}(\phi)&=& \frac{\beta NN' c_{1}}{\omega} \cos(\omega \phi) -
   \frac{\beta N N' c_{2}}{\omega} \sin(\omega \phi) +c_{3}   \mbox{,} 
 \\
A^{1}(\phi)&=&c_{1} \sin(\omega \phi)+c_{2}\cos(\omega \phi) \mbox{,}
\\
A^{2}(\phi)&=& \frac{N(N^{\phi}\beta +1)c_{1}}{\omega} \cos(\omega \phi) -
\frac{N(N^{\phi}\beta +1)c_{2}}{\omega}  \sin(\omega \phi) +c_{4}  \mbox{.}
\end{eqnarray}
with the `angular frequency' $\omega$ given by
\begin{equation}
\omega^{2} =N^{2}(N^{\phi}\beta+1)^{2}-\beta^{2}N^{2}N'^{2}.
\end{equation}
$c_{1},c_{2},c_{3},c_{4}$ are constants of
integration. The constants $c_{3}$ and $c_{4}$ are not independent.
Substituting the solutions back into the differential equations
system we see that: $c_{4}=(R\alpha^{2} \beta /\sqrt{-8M+R^{2}\alpha^{2}}) c_{3}$.
We can also  analize the \underline{orbit of constant time}.
The components of vector ${\bf A}$ satisfy
\begin{eqnarray}
\frac{dA^{1}}{d\phi}-N A^{2}&=& 0  \mbox{,}   \\
\frac{dA^{2}}{d\phi}+N A^{1}&=& 0    \mbox{.}
\end{eqnarray}
We assume that the orbits begin in $\phi=0$ and that after $n$ laps, we have
$\phi=2\pi n$. We also assume  that $A^{0}= \gamma$ and $A^{3}=$constant. 
The other components are easily integrated to
\begin{eqnarray}
         A^{1}(2\pi n)&=& c_{1}\sin(N2\pi n)+c_{2}\cos(N 2\pi n) \mbox{,}       \\
         A^{2}(2\pi n)&=& c_{1}\cos(N2\pi n)-c_{2}\sin(N 2\pi n)  \mbox{.}
\end{eqnarray}
For finite $r$ there is a deficit angle when the vector is  parallel
transported by $2\pi$, except when $n\sqrt{-8M+\alpha^{2}r^{2}+\frac{16J^{2}}{r^{2}}}$
equals an integer. The quantization condition of the orbits for 
holonomy invariance implies that
\begin{equation}
R^{2}=\frac{1}{2\alpha^{2}}\left(8M+(m/n)^{2}\pm \sqrt{(8M+(m/n)^{2})^{2}-(8\alpha J)^{2}}  \right),
\end{equation}
where $m$ is a non-zero integer. The above expression shows that for
geodesic motion along curves with constant time in the background of
a black string the holonomy invariance occurs after $n$ loops.

Another kind of curve to be studied are the
\underline{time-like circles}, whose tangent vector
is given by $X^{a}=\left(N \beta,0,r(N^{\phi}\beta+1),0\right)$. For
time-like curves the $N^{2}\beta^{2} - r^{2}(N^{\phi}\beta+1)^{2}<0$
condition is required. After $n$ loops we get
\begin{eqnarray}
\triangle A^{0}(2\pi n)&=& \frac{\beta N N' c_{1}}{\omega} \left[\cos(2\pi n \omega ) - 1\right]-
\frac{\beta N N' c_{2}}{\omega} \sin(2\pi n \omega )\mbox{,}
        \\
\triangle A^{1}(2\pi n)&=&c_{1} \sin(2\pi n \omega)-c_{2}\left[\cos(2\pi n \omega) -1\right]
 \mbox{,}\\
\triangle A^{2}(2\pi n)&=& \frac{N(N^{\phi}\beta+1) c_{1}}{\omega} \left[\cos(2\pi n\omega) - 1\right]-
\frac{N(N^{\phi}\beta+1) c_{2}}{\omega} \sin(2\pi n \omega )   \mbox{,}
\end{eqnarray}
where, $\triangle$ stands for the difference in the
components of the tangent vector before and after being transported.

Other very interesting paths are the \underline{radial paths}. In this 
case we can take the tangent vector as $X^{\mu}=[X^{t},X^{r},0,0]$,
so we can write for the tetrad components
\begin{equation}
X^{a}=\left(NX^{t},N^{-1}X^{r},0,0\right),
\end{equation}
For simplicity, we choose radial null geodesics. This condition
is satisfied if and only if $N^{2}(X^{t})^{2}=N^{-2}(X^{r})^{2}$.
Setting the curve parameter as $r$, the set of differential equations
(\ref{comp1}) and (\ref{comp2})  reduces to
\begin{eqnarray}
\frac{dA^{0}}{dr}+N'N^{-1}A^{1}&=&0
\mbox{,}
\\
\frac{dA^{1}}{dr}+ N'N^{-1}  A^{0}&=&0
\mbox{.}
\end{eqnarray}
The general solution for this system is given by
\label{}
\begin{eqnarray}
A^{0}=c_{1}N^{-1}+c_{2}N
\mbox{,}
\\
A^{1}=c_{1}N^{-1}-c_{2}N
\mbox{,}
\end{eqnarray}
where $c_{1}$ and $c_{2}$ are integration constants. Fixing
$c_{1}=1$ and $c_{2}=0$ by imposing that $A^{0}=X^{0}$
and $A^{1}=X^{1}$ we obtain that
\begin{equation}
A^{0}=A^{1}=\left( -8M+\alpha^{2}r^{2}+16J^{2}/r^{2}   \right)^{-\frac{1}{2}}.
\end{equation}
Using $t$ as a parameter and turning $r,\phi,z$ as a constant,
we can solve the equations (\ref{comp1}) and (\ref{comp2}), yielding the
following solutions
\begin{eqnarray}
A^{0}&=&A^{0}(0)\cosh (N N' t) + A^{1}(0)\sinh(N N' t)\mbox{,}\\
A^{1}&=&A^{1}(0)\cosh (N N' t) -A^{0}(0)\sinh(N N' t) \mbox{,}
\end{eqnarray}
with $A^{2}$ and $A^{3}$ are constants.


\section{Spinorial Parallel Transport}

In this section we are  interested in the study of parallel transport
 of spinor in the geometry of the rotating black string. 
An important question that emerges when we study
the parallel transport of vectors is what happens
with more complex fields when they also undergo a parallel
transport. We know that vectors in a curved space-time
may be changed after a complete loop if there is a non-vanishing
curvature in the region surrounded by the loop.
So, we are interested
in understanding what comes about when spinors
are parallel transported in the geometry of a black string.
Spinors provide a linear representation of the group
of rotations in $n$-dimensional space.

The  covariant differentiation $\nabla_{\mu}$ of a spinor is defined
in terms of the spinor connection, $ \Gamma_{\mu}$,
\begin{equation}
\label{co_dev}
\nabla_{\mu}\psi=\partial_{\mu}\psi -\Gamma_{\mu}\psi.
\end{equation}
The spinor connection is given in terms of the Dirac matrices
$\gamma_{\alpha}$,
\begin{equation}
\Gamma_{\mu}=-\frac{1}{4}\omega^{\alpha}_{\nu \mu}\gamma_{\alpha}\gamma^{\nu}.
\end{equation}

For orbits with constant time we found the following expression
for the spin connection in terms of the Pauli´s spin matrices
\begin{equation}
\Gamma_{\phi}d\phi=-\frac{1}{2}N i \sigma_{3}d\phi.
\end{equation}
The phase associated with the above spinor connection is 
given by
\begin{eqnarray}
U(\gamma)&=&\exp\left(-\oint \Gamma_{\phi}d\phi \right)\nonumber \\
        &=& \exp \left(-2\pi \Gamma_{\phi}\right)\mbox{,}
\end{eqnarray}
expanding this term we obtain
\begin{equation}
U(\gamma)=\cos(\pi N)-i \sigma_{3}\sin(\pi N).
\end{equation}
For orbits with constant azimuthal angle, the unique
non-null term is 
\begin{equation}
\Gamma_{t}=\frac{1}{2}N\frac{\partial N}{\partial r}\sigma^{1}-
\frac{1}{2}N N^{\phi}i \sigma^{3}.
\end{equation}
The expression for the holonomy is obtained in a similar way 
of previous cases, and the matrix of temporal holonomy is given by
\begin{eqnarray}
  \label{spintemp}
  U(\gamma)= &1&-\frac{\Gamma_{t}}{[ N^2( \frac{1}{2}[ N^{\phi}-{ N'}^{2}])]^{\frac{1}{2}}}\sinh\{\sqrt{N^2( \frac{1}{2}[N^{\phi}- {N'}^{2}])}T\} \\ \nonumber &+& \frac{\Gamma_{t}^{2}}{[N^2( \frac{1}{2}[N^{\phi}- {N'}^{2}])]}\left[\cosh\{\sqrt{N^2( \frac{1}{2}[N^{\phi}- {N'}^{2}])}T\}-1\right].
\end{eqnarray}
 
For a non-rotating black string we have the following holonomy
matrix
\begin{equation}
U(\gamma)= 1- \frac{\Gamma_{t}}{r\alpha^2/2}\sinh(r\alpha^2T/2)+
\frac{\Gamma_{t}^{2}}{r^{2}\alpha^4/4}\left[\cosh(r \alpha^2T/2)-1\right],
\end{equation}
which leads to the following matrix representation
\begin{eqnarray}
U(\gamma)=
\left(
\begin{array}{cc}
        \cosh r \alpha^{2}T/2 & -\sinh r \alpha^{2}T/2 \\
        -\sinh r \alpha^{2}T/2 & \cosh r \alpha^{2}T/2 \\
\end{array}
\right)
\mbox{.}
\end{eqnarray}
It follows that
\begin{equation}
W(\gamma)=2\cosh(r\alpha^{2}T/2).
\end{equation}
Finally we investigate the most general paths in spinor parallel 
transport in the background of the black string. 
 The spinorial connection is considered with 
$\Gamma_{s}=\beta \Gamma_{t} + \Gamma_{\phi}$ where the connection is given by
\begin{eqnarray}
\label{conesp}
\Gamma_{s}=\frac{1}{2}\beta N N' \sigma^{1} -\frac{i}{2}N(N^{\phi}+1)\sigma^{3}
\end{eqnarray}
this connection has the following properties $\Gamma_{s}^{3}=-A^{2}\Gamma_{s}$ and $\Gamma_{s}^{4}=A^{2}\Gamma_{s}^{2}$ were $A^{2}=\frac{1}{2}[(N^{2}(\beta N^{\phi} + 1)^{2}- \beta^{2}(N N')^{2}]$.

The holonomy associated with this path is given by
\begin{eqnarray}
  \label{genespin}
  U(\gamma)= e^{\{-\int_{0}^{2 \pi n} \Gamma_{s}d\phi\}}
\end{eqnarray}
We can write the following expression for the holonomy  matrix of

\begin{eqnarray}
  \label{matrixsping}
U(\gamma)= 1 + \frac{2\Gamma_{s}}{\omega}\sin\{ \pi n \omega  \}  
+ \frac{(2 \Gamma_{s})^{2}}{{\omega}^{2}}[\cos\{ \pi n \omega \} - 1] ,
\end{eqnarray}

were $\omega^{2}= [N^{2}(\beta N^{\phi} +1) -\beta^{2}(NN')^{2}]$. Note that $\omega$ is the same that in the case of general paths in vector transport. The same analysis can be done in this case for a class of special orbits.

We return to the analysis circular orbits.
Instead of vectors, we will study the parallel transport 
of spinors. In order to do this,
we will make the covariant derivative (\ref{co_dev}) null
\begin{equation}
\partial_{\mu}\psi -\Gamma_{\mu}\psi=0.
\end{equation}
The first case to be analyzed will be of orbits with 
constant time, in this case the unique non-null
spinorial connection is $\Gamma_{\phi}$,
thus, we will have the following system to solve
\begin{eqnarray}
\partial_{\phi}\psi_{1}-\frac{1}{2}N i \psi_{1}&=&0, \\
\partial_{\phi}\psi_{2}+\frac{1}{2}N i \psi_{2}&=&0 .
\end{eqnarray}
These equations are easily integrated to
\begin{eqnarray}
\psi_{1}&=&\psi_{1}(0)\exp\left(\frac{iN\phi}{2}\right), \\
\psi_{2}&=&\psi_{2}(0)\exp\left(-\frac{iN\phi}{2}\right).
\end{eqnarray}
This result shows that when the spinor is parallel transported
its components are changed from $\psi(0)$ to $\psi$.
There are certain values of the radius of the orbit
that preserve the spinor's components,
this means that there is no deficit angle.
After $n$ complete loops the square of the critical radius
is given by
\begin{equation}
r^{2}=\frac{[(2m/n)^{2}+8M]\pm
\sqrt{[(2m/n)^{2}+8M]^{2}-64J^{2}\alpha^{2}}}{2\alpha^{2}},
\end{equation}
where $m$ is integer.
This result shows that combinations of certain values of the
mass, curvature  and the angular moment can make the holonomy null.


\section{Concluding Remarks}


In this article we evaluate the loop variables associated with
the black string metric. We use the gravitational loop variable and
the holonomy to characterize the space-time of the black string.
We have shown by explicit computation from the metric corresponding to 
a rotating black string that the loop variables are a combination of 
rotations and boosts in
this spacetime. The holonomy transformations are elements of the 
$ R\times SO(2,2)$ group. We can see from the results an effect, similar 
to the gravitational Aharanov-Bohm effect in the cosmic string case, 
when vectors are 
transported in circles around the black string. The very important 
difference is that in the cosmic string case the spacetime outside 
the defect is flat, in the present case the exterior region of the 
defect has constant curvature. This effect is equivalent to the 
effects that appear in parallel transport of vectors or spinors in 
the background of the cosmic string with internal structure. 

We remark that there exist studies in the literature \cite{prd:can,plb:witt} involving Wilson loops 
and BTZ black holes, but in a very different context from ours. These works use Wilson loops, 
in  2+1 Chern-Simons gravity, as a means to obtain the BTZ metric. In our work we already 
have the metric and use the  loop variables as a means of geometric characterization of the 
black string spacetime. Since we had an extra linear dimension added to the BTZ space in order 
to obtain the black string, our results can be used to describe paralell transport of vectors or 
spinors in the BTZ background. The only effect will be a dimensional reduction of the matrices 
describing the holonomy transformations.

Due to the overall curvature of the black string spacetime, one would expect 
the holonomy to be nonzero everywhere. Nevertheless, there are special regions, specified 
by some values of the radial coordinate, which do present vanishing holonomy.
This is the  band of holonomy invariance \cite{cqg:rem} for the loop variables in this 
spacetime that implies in holonomy quantization. This effect is demonstrated here for a spacetime 
that is not asymptotically flat, but radially asymptotically anti-de 
Sitter. We analyzed also the loop variables for the spinorial case and 
demonstrated the same band holonomy properties observed by Rothman, 
Ellis and Murugan\cite{cqg:rem}.  We demonstrated, for the vectorial case, 
that loop variables satisfy the Mandelstam relations and that the 
Wilson loop presented the same  band of holonomy invariance. Finally we remark 
the importance of the analysis of loop variables in the spacetime 
that are asymptotically anti-de Sitter as a previous analysis of the 
global properties of the black string in a D-brane\cite{jhep:emp1,jhep:emp2,prd:haw} scenario.     
\newline

{\bf Aknowledgement} \quad We thank to CAPES (PROCAD), CNPq and PRONEX for
financial support.

\end{document}